\documentclass[aps,pra,showpacs,twocolumn]{revtex4}
\usepackage{eurosym}
\usepackage{amsfonts}
\usepackage{amsmath}
\usepackage{amssymb}
\usepackage{graphicx}
\usepackage{epsfig}
\usepackage{color}

\providecommand{\U}[1]{\protect\rule{.1in}{.1in}}

\begin{document}

\preprint{}
\title{High-order harmonic generation by enhanced plasmonic near-fields in metal nanoparticules}

\author{T. Shaaran$^{1}$}
\author{M. F. Ciappina$^{1,2}$}
\author{R. Guichard$^{3}$}
\author{J. A. P\'erez-Hern\'andez$^{4}$}
\author{M. Arnold $^{5}$}
\author{T. Siegel $^{5}$}
\author{A. Za\"ir $^{5}$}
\author{M. Lewenstein$^{1,6}$}

\affiliation{$^{1}$ICFO-Institut de Ci\`encies Fot\`oniques, Mediterranean Technology
Park, 08860 Castelldefels (Barcelona), Spain}
\affiliation{$^{2}$Department of Physics, Auburn University, Auburn, Alabama 36849, USA}
\affiliation{$^{3}$Laboratoire Chimie-Physique, Mati\`ere et Rayonnement, Universit\`e Pierre et Marie Curie UMR 7614, F-75231 Paris 05, France}
\affiliation{$^{4}$Centro de L\`aseres Pulsados (CLPU), Parque Cient\`ifico, 37185 Villamayor, Salamanca, Spain}
\affiliation{$^{5}$Imperial College London, Department of Physics, Blackett Laboratory Laser Consortium, London SW7 2AZ, United Kingdom}
\affiliation{$^{6}$ICREA-Instituci\'o Catalana de Recerca i Estudis Avan\c{c}ats, Lluis
Companys 23, 08010 Barcelona, Spain}


\begin{abstract}
We present theoretical investigations of high-order harmonic generation (HHG) resulting from the interaction of noble gases with localized surface plasmons. These plasmonic fields are produced when a metal nanoparticle is subject to a few-cycle laser pulse. The enhanced field, which largely depends on the geometrical shape of the metallic structure, has a strong spatial dependency. We demonstrate that the strong non-homogeneity of this laser field plays an important role in the HHG process and leads to a significant increase of the harmonic cut-off energy. In order to understand and characterize this new feature, we include the functional form of the laser electric field obtained from recent attosecond streaking experiments [F. S{\"u}{\ss}mann and M. F. Kling, Proc. of SPIE, {\bf Vol. 8096}, 80961C (2011)] in the time dependent Schr\"odinger equation (TDSE). By performing classical simulations of the HHG process we show consistency between them and the quantum mechanical predictions. These allow us to understand the origin of the extended harmonic spectra as a selection of particular trajectory sets. The use of metal nanoparticles shall pave a completely new way of generating coherent XUV light with a laser field which characteristics can be synthesized locally.
\end{abstract}

\maketitle

When matter, i.e. atoms or molecules, is exposed to short and intense laser radiation, non-linear phenomena are triggered as a consequence of this interaction. Amongst these phenomena, high-order harmonics generation (HHG) process~\cite{McPherson1987,Huillier1991} has attracted considerable interests, since it is one of the most reliable pathways to generate coherent light from the ultraviolet (UV) to extreme ultraviolet (XUV) spectral range. As a result, HHG has proven to be a robust source for the generation of a PHz attosecond pulses train~\cite{corkumnat}, that can be temporally confined to a single XUV attosecond pulse, now with kHz repetition rates~\cite{Scrinzi2004}. Thanks to its remarkable properties, HHG can be used as well to extract temporal and spatial information with both attosecond and sub-{\AA}ngstr\"om resolution on the generating system~\cite{manfred_rev}. Hence, HHG represents a considerable tool to enable scrutinizing the atomic world with its natural temporal and spatial scales~\cite{pacer,rabitt,olga1,olga2,mairesse,power}.

The intuitive physical mechanism behind HHG, for a single atom or molecule (referred to as 'single emitter'), has been well established in the so-called three steps or simple man's model~\cite{corkum,sfa,kulander}: in the first step, an electronic wave packet is released the continuum by tunnel ionization through the potential barrier as a consequence of the non-perturbative interaction of the single emitter with the laser field. In the second step, the emitted electronic wave packet propagates away from its ionic core in the continuum to be finally driven back when the laser electric field changes its sign. In the final step, upon its return, the electronic wave packet may recombine with the core and the system relaxes the excess kinetic energy acquired by radiating a high-harmonic photon.

In order to experimentally control the high harmonic features two main types of approaches have been attempted. The first one is based on the control of the HHG process via the manipulation of the laser field characteristics in time and/or space~\cite{joseprl2013}. The second one is based on the control of
macroscopic properties of the target samples (i.e. phase matching) leading to very ingenious target geometries~\cite{constant,tenio1,tenio2}. However,
all these approaches rely on detecting the far-field properties of the harmonic yield which is a consequence of collective single emitter's ones. Therefore, it is legitimate to study how a synthesized single emitter could lead to new parameters for HHG control. Metal nanoparticles are a matter of choice since the spatial geometry of the nanoparticle and the material used can be chosen to confer a spatial transverse non-homogeneity to the laser field. One of the first demonstrations of such an effect was obtained for surface plasmonic resonances that can locally amplify the laser field~\cite{kim}. The local electric fields resulting from such resonances can reach an enhancement greater than 20 dB~\cite{muhl, schuck}. Consequently, when a femtosecond low intensity laser pulse couples to the plasmonic mode of the metal nanoparticle, it initiates a collective oscillation among free charges (essentially electrons) within the metal. A location of highly amplified electric field is thus created while these free charges redistribute this field around the metal nanostructure. The enhanced field is well above the threshold for generating high harmonics. So, by injecting noble gases surrounding the nanoparticle, HHG can be produced. Particularly, whilst using gold bow-tie shaped nanostructures, it has been demonstrated that the initial laser field (800 nm laser with intensity $10^{11}$ W/cm$^{2}$) can be enhanced sufficiently to generate XUV photons, i.e. from the 7th (114 nm) to the 21st (38 nm) harmonic order. Furthermore, the high harmonics radiation generated from enhanced laser field, localized at each nanostructure, acts as a point-like source, enabling collimation or focusing of this coherent radiation by means of constructive interference. This localization of the enhancement confers to the transverse laser field profile its non-homogeneity in the region where the electron dynamics is taking place. Besides, the interaction length along the propagation direction of the laser field is restricted to few nanometers so that no phase conditions needs to be considered in order to observe and calculate the emitted harmonic yield.  As a consequence, spatially arranged nanostructures open a wide range of possibilities to enhance or shape the spectral and spatial properties of the fundamental laser field and the harmonic field~\cite{kim}. These two features imply strong modifications in the harmonic spectra which will raise the interest of the strong field community to utilize nanoparticles surrounded by gas atoms or molecules as a new type of target~\cite{husakou,yavuz,ciappi2012,tahir2012}.
However, the initial thrill about the utilization of plasmonic fields for HHG in the XUV range, was put in debate by recent findings~\cite{ropersnat, Kimreply, corkum_priv}. Fortunately, alternative ways to amplify coherent light using plasmons were explored. To cite only a couple of prominent examples, the production of high energy photoelectrons using plasmonic near enhanced fields from dielectric nanoparticles~\cite{klingnature}, metal nanoparticles~\cite{klingspie,klingprb,lastkling} and metal nanotips~\cite{herink,peter2006,peter2010,peter2011,peter2012} appears to be perfectly plausible. Besides the question of damage threshold of such nanotargets has been highlighted and indicates that new routes using initially low intensity laser field need to be considered~\cite{funnel}.

In this contribution we investigate how HHG yield can be controlled using enhanced near-fields. These fields are obtained only when a metal spherical nanoparticle is illuminated by a few-cycle laser field . This particularly ultra-short interaction regime confers to these near-fields a strong spatial dependence that can be accessible experimentally using attosecond streaking techniques~\cite{garching}. The plasmonic enhanced near-field in the vicinity of these nanotargets could act as a femtosecond coherent source and therefore drive strong laser-matter processes in the surrounding atomic or molecular gas , e.g. high-order harmonic generation. In addition, the enhancement of the fundamental laser field reaching one or more orders of magnitude, depending on the size and material employed, would allows us to consider low input laser intensity and thus to work well below the damage threshold of the nanotarget. Using both quantum mechanical and classical approaches, we predict that the signature of such synthesized fields is pronounced in the harmonics cutoffs which now extend far beyond the conventional semiclassical limits. 

Most of the numerical and semiclassical approaches to study high-order harmonic generation (HHG) are largely based on the assumption that the laser electric field ($\mathbf{E}(\mathbf{r},t)$) and its associated vector potential ($\mathbf{A}(\mathbf{r},t)$) are spatially homogeneous in the region where the electron dynamics takes place, i.e. $\mathbf{E}(\mathbf{r},t)=\mathbf{E}(t)$ and $\mathbf{A}(\mathbf{r},t)=\mathbf{A}(t)$~\cite{keitel,krausz}. Nonetheless, near-fields generated in the vicinity of metal nanoparticles are not spatially homogeneous and we address the question on how such non-homogeneity could be revealed in the harmonic spectra. In the case of homogeneous fields, the HHG process has been theoretically tackled using different approaches (for a summary see e.g.~\cite{book1,book2} and references therein). In this article, we compute the HHG spectra by including the spatial dependence of the field in the dimensionally reduced Time Dependent Schr\"{o}dinger Equation (TDSE) by considering the actual functional form of the laser electric field spatial dependence, obtained from attosecond streaking experiments. The TDSE in one spatial dimension and for a model atom can be written as~\cite{keitel}:
\begin{eqnarray}
\mathrm{i}\frac{\partial \Psi (x,t)}{\partial t} &=&\mathcal{H}(t)\Psi (x,t)
\label{tdse} \\
&=&\left[ -\frac{1}{2}\frac{\partial ^{2}}{\partial x^{2}}%
+V_{a}(x)+V_{l}(x,t)\right] \Psi (x,t).  \notag
\end{eqnarray}%
In here, $V_{a}(x)$ is the atomic potential and $V_{l}(x,t)$ represents the potential due to the laser electric field. For the $V_{a}(x)$ , we use a soft-core potential to avoid singularities.
\begin{equation}
\label{atom}
V_{a}(x)=-\frac{1}{\sqrt{x^{2}+a^{2}}}.
\end{equation}%

This potential was first proposed in~\cite{eberly} and has been widely used in studies of laser-matter processes in atoms. Certainly the model potential is not suitable to predict the actual atomic structural information present in the harmonic spectrum, but it is completely applicable to characterize the HHG cutoff, once the ionization potential of a given atom is set. A particular ionization potential can be defined by varying the parameter $a$ in Eq. (\ref{atom}). In this work we use $a=1.62$ to model Xenon atoms with $I_{p}=12.1299$ eV ($0.446$ a.u.). In addition, we assume that the xenon atom is in its ground state before the laser field ($t=-\infty $) is turned on. Equation (\ref{tdse}) is solved numerically by using the Crank-Nicolson method~\cite{keitel}. To avoid spurious reflections from the spatial boundaries, at each time step, the electron wave function is multiplied by a mask function~\cite{mask}. The potential due to the linearly polarized laser electric field in the $x$-axis, $E(x,t)$ in (\ref{tdse}), is given by:
\begin{equation}
V_{l}(x,t)=-E(x,t)\,x.
\end{equation}%
We employ  the function given by~\cite{klingspie} to define  $E(x,t)$, i.e.
\begin{equation}
E(x,t)=E_{0}\,f(t)\,\exp (-x/\chi )\sin (\omega t+\phi ).  \label{electric}
\end{equation}%

In Eq. (\ref{electric}), $E_{0}$, $\omega $, $f(t)$ and $\phi $ are the peak amplitude, the laser field frequency , the field envelope and the carrier envelope phase (CEP), respectively. The spatial dependence of the plasmonic near-field is given by the term $\exp(-x/\chi )$ and it depends on both the size and the material of the spherical nanoparticle used. Eq. (\ref{electric}) is valid for $x\ge R_0$, where $R_0$ is the radius of the metal nanoparticle, i.e. for values of $x$ outside of the metal nanoparticle. Additionally, it is important to note that the electron motion stands in the region $x\ge R_0$ with $(x+R_0)\gg0$. In this work we consider the laser field having a sin$^{2}$ envelope to avoid DC components:
\begin{equation}
f(t)=\sin ^{2}\left( \frac{\omega t}{2n_{p}}\right)   \label{ft}
\end{equation}%
where $n_{p}$ is the total number of optical cycles so that the total pulse duration is $\tau =2\pi n_{p}/\omega $.

The harmonic yield of the atom is obtained by Fourier transforming the acceleration $a(t)$ of the electronic wave packet~\cite{schafer}:
\begin{equation}
D(\omega )=\left\vert \frac{1}{\tau }\frac{1}{\omega ^{2}}\int_{-\infty
}^{\infty }\mathrm{d}t\mathrm{e}^{-\mathrm{i}\omega t}a(t)\right\vert ^{2}
\end{equation}%
with $a(t)$ obtained by using the commutator relation:
\begin{equation}
a(t)=\frac{\mathrm{d}^{2}\langle x\rangle }{\mathrm{d}t^{2}}=-\langle \Psi
(t)|\left[ \mathcal{H}(t),\left[ \mathcal{H}(t),x\right] \right] |\Psi
(t)\rangle.
\label{accel1D}
\end{equation}%
In here, $\mathcal{H}(t)$ and $\Psi (x,t)$ are the Hamiltonian and the electron wave function defined in Eq. (\ref{tdse}), respectively. The function $D(\omega )$ is known as the dipole spectrum, which gives rise to the spectral profile measured in HHG experiments.
\begin{figure}[htb]
\centering
\includegraphics[width=0.3\textwidth]{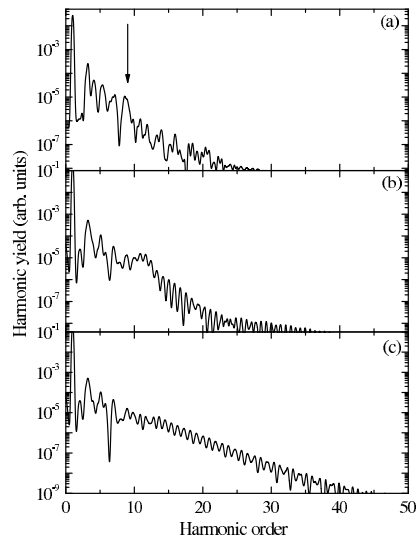}
\caption{High-order harmonic generation (HHG) spectra for Xe (ionization potential $I_{p}=-0.446$ a.u.), laser wavelength $\protect\lambda=720$ nm
and intensity $I=2\times10^{13}$ W$\cdot$cm$^{-2}$. We use a sin$^{2}$ shaped pulse with $n=5$ (about 13 fs of total time and $\approx6.5$ fs FWHM). Panel (a) shows the homogeneous case,
panel (b) $\protect\chi=50$ and panel (c) $\protect\chi=40$. The arrow in panel (a) indicates the cutoff predicted by the semiclassical model~\protect\cite{sfa}. }
\label{fig:figure1}
\end{figure}

Figure 1 depicts the harmonic spectra for Xenon generated by a laser pulse with $I=2\times 10^{13}$ W/cm$^{2}$, wavelength $\lambda =720$ nm and a total pulse duration of 13 fs, i.e $n_{p}=5$ (this pulse corresponds to $\approx6.5$ fs FWHM). We have chosen these parameters to perform calculations consistently with experimental conditions ~\cite{klingspie}. For the case of a homogeneous field, with this low fundamental laser intensity, no harmonic beyond the $9^{th}$ order are observed (see the arrow in panel a). The spatial decay constant $\chi$ quantifies the non-homogeneity due to the nanoparticle. It varies together with the size of the particle and the metal employed. Changing $\chi$ is therefore equivalent to choosing the type of nanoparticle used, which leads to overcome the semiclassically predicted cutoff limit and reach higher harmonic orders. For instance with $\chi =40$, harmonics in the mid 20s are obtainable (panel c) and with $\chi =50$ we observe harmonic orders well above the $9^{th}$ (a clear cutoff at $n\approx 15$ is visible) (panel b). In addition to this cut-off extension, we observe a change in the harmonic periodicity. This is related to the breaking of symmetry imposed by the induced non-homogeneity of laser electric field (see. Eq.~(\ref{electric})).
\begin{figure}[htb]
\centering
\includegraphics[width=0.4\textwidth]{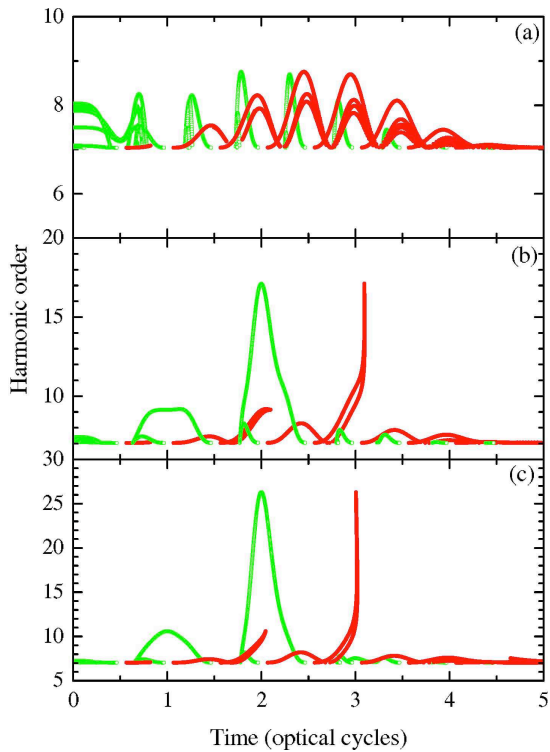}
\caption{(Color online) Total energy of the electron (expressed in harmonic order) driven by the laser field calculated from Newton's second law and plotted as a function of the ionization time (green circles) or the recollision time (red circles). (a) Homogeneous case, (b) $\protect\chi=50$ and (c) $%
\protect\chi=40$.}
\label{fig:figure2}
\end{figure}

In order to understand the relationship between the laser field induced non-homogeneity and the harmonic cutoff extension, we consider the semi-classical three-step model~\cite{corkum,sfa}. As it was already pointed out in~\cite{ciappi2012,ciappi_opt}, this new feature may appear due to the combination of two main factors, namely the non-homogeneous character of the laser electric field and the confinement of the electronic motion. 
For an homogeneous field, it is well established that the position of the high-order harmonic cutoff holds as:
\begin{equation}
n_{c}=(3.17U_{p}+I_{p})/\omega   \label{cutoff}
\end{equation}%
where $n_{c}$ is the harmonic order at the cutoff, $\omega $ the laser frequency, $U_{p}$ the ponderomotive energy (defined by $U_{p}=I/4\omega^{2}
$, with $I$ being the laser intensity in a.u.) and $I_{p}$ the ionization potential of the atom or molecule.
In our case we solve numerically the Newton equation for an electron moving in a linearly polarized (in the $x$-axis) electric field with the same parameters
used in the 1D-TDSE calculations, i.e.  with a time and space dependent electric field $E(x,t)$ of the form Eq.~(\ref{electric}). Then we find the
numerical solution of:
\begin{eqnarray}
\ddot{x}(t) &=&-\nabla _{x}V_{laser}(x,t)  \notag  \label{newton} \\
&=&-E(x,t)-\left[ \nabla _{x}E(x,t)\right] x(t)  \notag \\
&=&-E(x,t)(1-\frac{x(t)}{\chi }),
\end{eqnarray}%
where $E(x,t)$ is defined by Eq. (\ref{electric}). Note that in (\ref{newton}), we have presented explicitly the time dependence of the electron trajectory, i.e.
$x=x(t)$. 
We consider initial conditions similar to the simple man's model: the electron starts at position zero at $t=t_{i}$ (the ionization time) with zero velocity, i.e. $x(t_{i})=0$ and $\dot{x}(t_{0})=0$. When the electric field reverses its direction, the electron returns to its initial position (i.e the electron \textit{recollides} or recombines with the parent ion) at a later time $t=t_{r}$ (the recollision time), i.e. $x(t_{r})=0$. The electron kinetic energy at the return time $t_{r}$ is calculated from: 
\begin{equation}
E_{k}(t_{r})=\frac{\dot{x}(t_{r})^{2}}{2}
\end{equation}%
and finding the value of $t_{r}$ (as a function of $t_{i}$) that maximizes this energy, fulfills Eq. (\ref{cutoff}).

Fixing the value of the ionization time $t_{i}$ it is possible to compute the classical trajectories and to numerically calculate the recollision times $t_{r}$, i.e. the $t_{r}$ when $x(t_{r})=0$. For a given ionization time $t_{i}$ the electron trajectory is completely determined, because the second order differential equation (\ref{newton}) and its initial conditions have a unique solution. In Fig.~2 panels (a)-(c), we show the dependence of the harmonic order upon the ionization time ($t_{i}$) and recollision time ($t_{r}$), calculated from $n=(E_{k}(t_{i,r})+I_{p})/ \omega $ with $I=2\times 10^{13}$ W cm$^{-2}$, $\lambda =720$ nm and a sin-squared shaped pulse of 5 cycles (total time duration 13 fs).

Panels (a), (b) and (c) depicts the cases of ($\chi\rightarrow\infty)$ (homogeneous field), $\chi =50$ and $\chi =40$ (two cases of non-homogeneous field), respectively. From panel (a), we observe that the maximum kinetic energy of the returning electron agrees with Eq. (\ref{cutoff}) (no harmonic order beyond $n_{c}\sim 9\omega$ is reached). On the other hand, panels (b) and (c) show how the non-homogeneity of the laser field modifies considerably the electron trajectories towards an extension of the harmonic cut-off energy. This is clearly present at $n_{c}\sim 18\omega$ (28 eV) and $n_{c}\sim 27\omega$ (42 eV) for $\chi =50$ and $\chi =40$, respectively. These last two cutoff values are indeed consistent with the quantum mechanical calculations presented in Fig.~1

In order to show how our model behaves for higher laser intensities we compute the harmonic spectra by increasing the intensity to $I=5\times
10^{13}$ W/cm$^{2}$, keeping all other parameters the same (we note that the saturation intensity of xenon is $\approx 8\times
10^{13}$ W/cm$^{2}$ and consequently our proposed values are well below this value). The results are shown in Fig.~3. In panels (b) and (c), the non-homogeneity of the laser field manifests itself as a clear harmonic cutoff extension, reaching values of $n_{c}\sim 60\omega$ (93 eV in energy) for $\chi=40$. These values show a highly nonlinear dependence of the harmonic cutoff with the spatial decay constant $\chi $. This non-linear behavior can be exploited to generate high order harmonic in the XUV regime using modest laser intensities.
\begin{figure}[htb]
\centering
\includegraphics[width=0.4\textwidth]{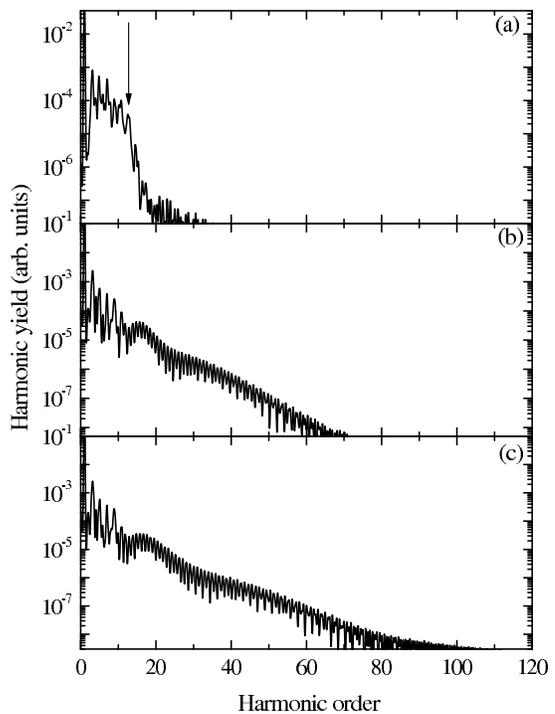}
\caption{Idem Fig. 1 but now the laser intensity is $I=5\times10^{13}$ W/cm$^{2}$.}
\label{fig:figure3}
\end{figure}
Finally, in Fig.~4, we plot the recollision time $t_{r}$ of the electron as a function of the ionization time $t_{i}$ as for the case presented in Fig.~1. For the ionization times confined between $0.5<t_t<1.5$ optical cycles, the long trajectories are those with recollision times $t_{r}\gtrsim 2.25$ optical cycles and they are visible only for the homogeneous case (red squares ($\color{red}\blacksquare $)) and for $\chi =50$ (green filled circles ($\color{green}\bullet $)). On the other hand, short trajectories are characterized by $t_{r}\lesssim 2.25$ optical cycles and these are present for both the homogeneous and non-homogeneous cases. Our results are consistent with those presented in~\cite{yavuz,ciappi2012,ciappi_opt}, although in our work we use a different functional form of the electric field resulting from experimental results. In Fig.~4, we observe how the long trajectories are strongly modified by the non-homogeneity of the laser field. Indeed the \textit{homogeneous} long trajectories (red squares ($\color{red}\blacksquare $)) with ionization times $t_{i}$ around the 1.25 and 1.75 optical cycles \textit{converge} with the short one into a unique trajectory set ($\color{blue}\blacktriangle $). Moreover, the branch with $t_{i}\sim 1.75$, has now ionization times smaller than in the homogeneous case, hence, the propagation time of the electron in the continuum increases so that it can gain a higher amount of kinetic energy ~\cite{yavuz,ciappi2012,ciappi_opt}, as confirmed in the classical calculations presented in Fig.~2.

The non-homogeneity of the laser field annihilates the long trajectories and only short trajectories are now responsible for the harmonic spectrum. This behavior is also observed using a non-homogeneous field with spatial linear dependence (for details see~\cite{ciappi2012,tahir2012,ciappi_opt,ciappiati}). Considering that the electric field strength at the ionization time for short trajectories is higher than for the long ones, and taking into account that the ionization rate is a highly nonlinear function of this electric field~\cite{keldysh65,ammosov87}, long trajectories are much less efficient than the short ones. Accordingly, this fact explains why the harmonic spectrum, obtained with a non-homogeneous laser field that \textit{modified} the trajectories shows an extended cutoff.
\begin{figure}[htb]
\centering
\includegraphics[width=0.4\textwidth]{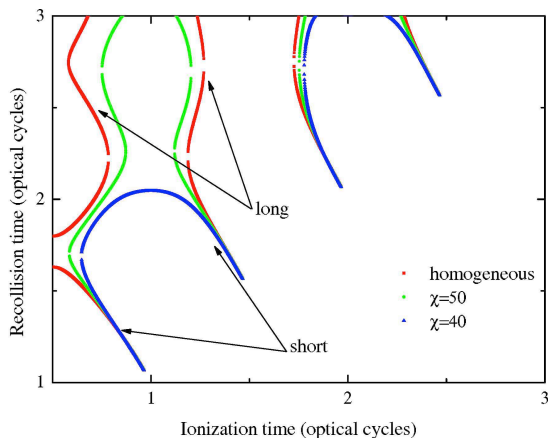}
\caption{(Color online) Dependence of the semi-classical trajectories on the
ionization ($t_i$) and recollision ($t_r$) times for different values of $\chi$. Red squares are homogeneous case, i.e. $\chi\rightarrow\infty$; green circles are $\protect\chi=50$ and blue triangles $\chi=40$.}
\label{fig4}
\end{figure}

In conclusion, we present how the high-order harmonic generation from xenon atoms is modified by using a plasmonic near enhanced field generated when a metal nanoparticle is illuminated by a short and intense laser pulse. The functional form of the resulting laser electric field is extracted from attosecond streaking experiments and incorporated in our quantum mechanical approach. We employ the numerical solution of the time dependent Schr\"odinger equation (TDSE) in reduced dimensions in order to calculate the harmonic spectra. We observe an extension in the harmonic cutoff position that could lead to the production of XUV coherent laser sources and opening the avenue to the generation of attosecond pulses from spatially inhomogeneous laser fields. This new feature is a consequence of the induced laser field non-homogeneity only, which modifies substantially the electron trajectories. Furthermore, our quantum mechanical numerical results agree with the classical simulations. A more pronounced increment in the harmonic cutoff, in addition with an appreciable growth in the conversion efficiency, could be reached varying, for instance, the radius of the spherical metal nanoparticles and by choosing the adequate metal materials. These new degrees of freedom could pave the way to enhance the harmonic spectra reaching the XUV regime with modest input laser intensities.

We acknowledge the financial support of the MICINN projects (FIS2008-00784
TOQATA, FIS2008-06368-C02-01 and FIS2010-12834); ERC Advanced Grant
QUAGATUA, Alexander von Humboldt Foundation and Hamburg Theory Prize (M.
L.). This research has been partially supported by Fundaci\'o Privada Cellex. J. A. P.-H. acknowledges support from Spanish
MINECO through the Consolider Program SAUUL
(CSD2007-00013) and research project FIS2009-09522,
from Junta de Castilla y Le\'on through the Program for
Groups of Excellence (GR27) and from the ERC Seventh
Framework Programme (LASERLAB-EUROPE, Grant
No. 228334)
A. Z. acknowledges the support from EPSRC Grant No. EP/J002348/1 and Royal Society Internal Exchanges 2012 Grant No. IE120539.
We thank Matthias Kling and Sergey Zherebstov for useful comments and suggestions.






\end{document}